\documentclass[aps,preprint,groupedaddress,superscriptaddress,longbibliography]{revtex4-1}

\usepackage{graphicx}% Include figure files
\usepackage{dcolumn}% Align table columns on a decimal point
\usepackage{bm}% bold math
\usepackage{multirow}
\usepackage{color}
\usepackage{ulem}

\begin{document}
%\linenumbers

\title{Many-body benchmarking of DFT local-registry energetics in bilayer InSe}

\author{Jeonghwan Ahn}
\affiliation{The Anthony J. Leggett Institute for Condensed Matter Theory, Department of Physics, University of Illinois at Urbana-Champaign, Urbana, IL 61801, USA
}
\author{Abdulgani Annaberdiyev}
\affiliation{Center for Nanophase Materials Sciences, Oak Ridge National Laboratory, Oak Ridge, TN 37830, USA}
\author{Jovan Nelson}
\affiliation{Applied Physics Program, Northwestern University, Evanston, Illinois 60208, USA}
\author{Nathaniel P. Stern}
\affiliation{Applied Physics Program, Northwestern University, Evanston, Illinois 60208, USA}
\affiliation{Department of Physics and Astronomy, Northwestern University, Evanston, Illinois 60208, USA}
\author{Hyeondeok Shin}%
\email{hshin@anl.gov}
\affiliation{Computational Science Division, Argonne National Laboratory, Argonne, Illinois 60439, USA}

\date{\today}

\begin{abstract}
{Density functional theory (DFT) is widely used to model twisted bilayers, but the accuracy of the local stacking energetics underlying such models remains uncertain. Here, we benchmark the local-registry landscape of bilayer InSe using diffusion quantum Monte Carlo (DMC). DFT predicts that AB, AAr, and ABr stackings, which share the same interfacial Se registry, are nearly degenerate within 1.5 meV/f.u. and exhibit nearly indistinguishable DFT charge-density responses. DMC instead separates these stackings by 8(5) and 41(4) meV/f.u., while the energy difference between the most stable and least stable registries reaches 60(7) meV/f.u.. These large energy separations show that the stacking energetics are not determined by the interfacial atomic motif alone but depend on the full registry and its associated many-body electronic response. More broadly, these results show that DFT-based moir\'e models can substantially underestimate local stacking-energy corrugation, with direct consequences for predicted structural relaxation, domain formation, and electronic reconstruction in twisted layered materials.
}
\end{abstract}

\maketitle

\section*{Introduction}
\label{sec:intro}

Layered materials provide a broad setting in which interlayer coupling controls structure, energetics, and electronic response. 
Even when the constituent monolayers are individually well understood, bringing two layers together introduces a new degree of freedom through the lateral registry between them.
Local stacking registry can reshape the equilibrium spacing, charge redistribution, and low-energy electronic structure~\cite{yoo2019atomic,carr2020electronic}. This registry dependence becomes more consequential in twisted bilayers, where the moir\'e pattern contains spatially varying local stacking environments. 
In small-angle structures, the moir\'e pattern can further develop local domains and out-of-plane corrugation, so different regions of the same twisted bilayer may sample different interlayer spacings as well as different lateral registries. 
The interlayer coupling of a twisted bilayer is therefore tied to the stacking-dependent bilayer landscape from which these local geometries are drawn.

Bilayer InSe is a particularly stringent system for addressing this issue because the local registries relevant to twisting are connected to experimentally realized interlayer arrangements. 
The $\beta$, $\epsilon$, and mixed-polymorph forms of InSe have been reported in bulk and few-layer samples~\cite{zhou2018inse,hilse2025mixed,song2026indium}, showing that different relative layer registries can occur within the same compound. 
These experimentally relevant registries provide natural local reference configurations for a twisted bilayer, where the lateral displacement between layers varies across the moir\'e pattern. 
Consequently, the twist problem in InSe inherits its local interlayer physics from the same high-symmetry stacking environments that define the untwisted bilayer landscape. 
A predictive description of twisted bilayer InSe therefore requires an accurate account of stacking-dependent interlayer coupling in the corresponding high-symmetry bilayer registries.

Density functional theory (DFT) has been the primary practical framework for studying InSe and related layered semiconductors, including their stacking-dependent and twist-dependent properties~\cite{mudd2013tuning,sun2018inse,yao2021electronic,pike2024understanding,magorrian2025strong}. 
This reliance on DFT is natural because the method can describe structural and electronic trends in large layered systems at an accessible computational cost. 
The difficulty is that interlayer coupling in layered materials is governed by a delicate interplay among short-range repulsion, nonlocal electronic correlation, and interlayer hybridization. 
Similar local interlayer spacings therefore do not guarantee similar binding energies, and a structurally reasonable geometry does not necessarily imply that the underlying interlayer energetics are described correctly. 
This issue is particularly important when nonlocal van der Waals attraction and registry-dependent interlayer hybridization act together, because their interplay controls the stacking-resolved interlayer potential and can depend sensitively on how exchange and correlation are treated within DFT.
In practice, a reliable description of bilayer and twisted InSe should resolve the stacking-dependent energy landscape and the associated electronic response on the same footing. 
These considerations motivate a many-body benchmark for the high-symmetry bilayer registries, not only to assess DFT at the bilayer level, but also to evaluate whether DFT provides an appropriate local-registry landscape for twisted bilayer InSe.

Here, diffusion quantum Monte Carlo (DMC)~\cite{foulkes01} reveals the stacking dependence hidden by the DFT description of bilayer InSe. 
At the DFT level, the sliding landscape is shallow and the band gap changes little with stacking. 
In particular, three stackings sharing the same interfacial Se registry remain nearly degenerate within 1.5 meV/f.u..
The many-body benchmark overturns this picture by splitting these stackings by up to 41(4) meV/f.u. and substantially enlarging the energy range between the most stable and least stable registries.
Density analysis supports this energetic hierarchy by revealing registry-dependent charge redistribution that is largely smoothed out in DFT. 
Since twisted bilayer InSe samples the same local stacking landscape, the DMC benchmark implies that the weak twist dependence suggested by DFT reflects a flattened local-registry landscape rather than the full many-body interlayer coupling.
More broadly, these results show that underestimating the local-registry energy scale can propagate into moir\'e models and affect predicted structural relaxation, domain formation, and electronic reconstruction in twisted layered materials.

\section*{Results and discussion}
\label{sec:results}

\subsection*{Density-functional local-registry landscape of bilayer InSe}

\begin{figure}
 \includegraphics[width=6.5in]{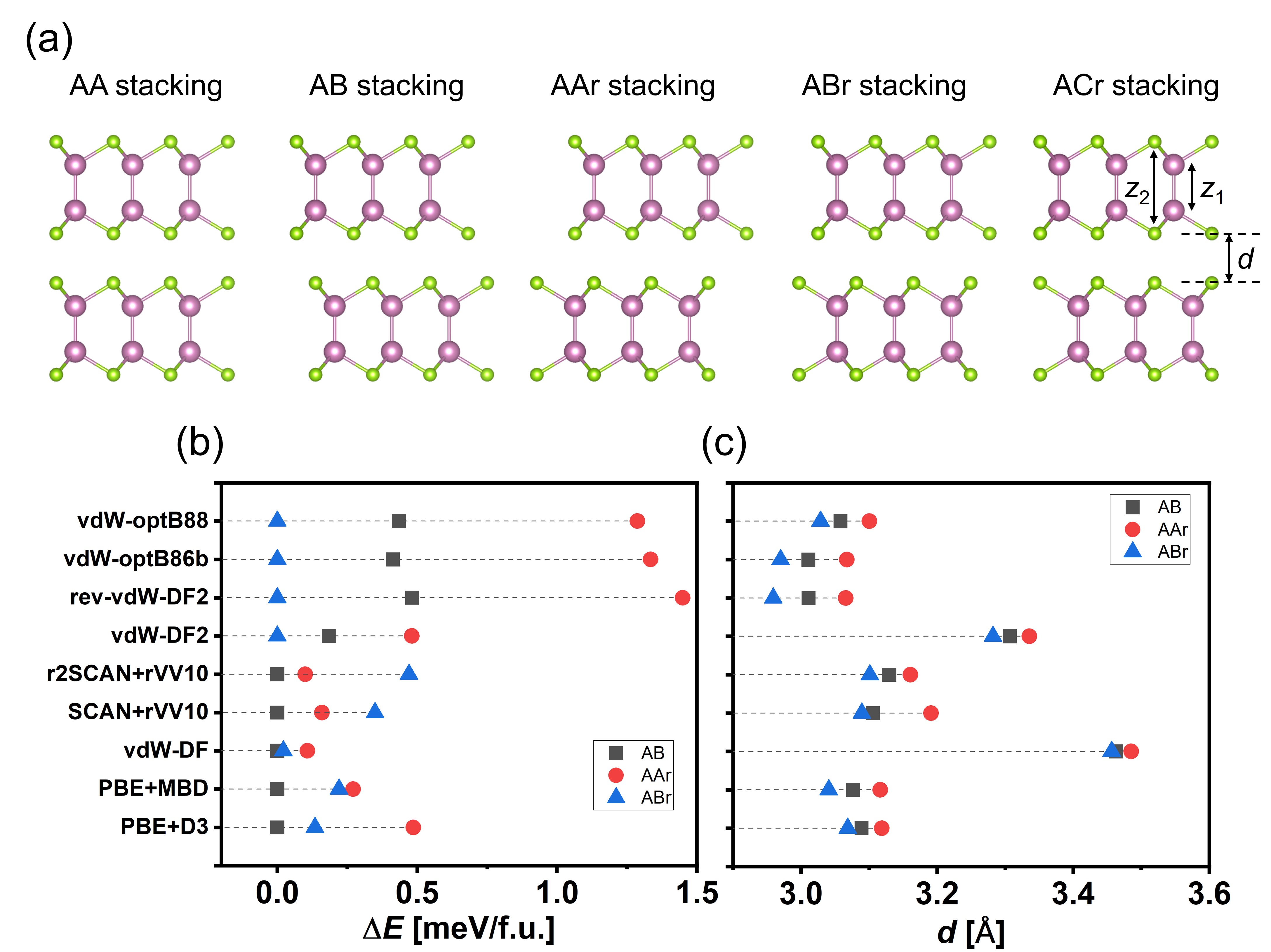}
 \caption{High-symmetry bilayer InSe registries and DFT benchmark quantities. (a) Atomic structures of the five high-symmetry stacking configurations considered in this work: AA, AB, AAr, ABr, and ACr. The interlayer distance $d$ is defined as the vertical separation between the two facing Se planes. (b) Relative DFT energies of the AB, AAr, and ABr stackings obtained with different exchange-correlation functionals. Energies are shown relative to the lowest-energy stacking for each functional. (c) Corresponding optimized interlayer distances. %\hscom{What does missing dotted line in r2SCAN+rVV10 and SCAN+rVV10 mean? Perhaps just mistake?}
 } 
 \label{fig:bilayer_DFT}
\end{figure} 

%\xtask{1. Fig.1: relative energetics and interlayer distances among stable stackings are important quantities needed to benchmark }

%\xtask{2. Fig.2: energy and interlayer-distance difference between stable and unstable ones are also important quantities because both stable and unstable simultaneously constitute twisted bilayer, which determines the energetics of twisted bilayer.}

We first establish the density-functional reference landscape for bilayer InSe. 
The five high-symmetry stackings considered in this work are shown in Fig.~\ref{fig:bilayer_DFT}: AA, AB, AAr, ABr, and ACr. 
Note that AC stacking is equivalent to AB by symmetry.
These registries are not merely isolated bilayer structures. 
They represent the local stacking environments that appear across the moir\'e pattern of twisted bilayer InSe. 
AB and AAr are particularly important because they correspond to the $\epsilon$-like and $\beta$-like layer registries, respectively. 
The remaining high-symmetry AA, ABr, and ACr configurations are also essential because a twisted bilayer samples both favorable and unfavorable local registries.
Its energetics is therefore determined by the full local-registry landscape rather than by the lowest-energy stacking alone.

Figure~\ref{fig:bilayer_DFT}(b) first compares AB, AAr, and ABr because these three stackings share the same interfacial Se arrangement.  
%\hscom{It should be AB, AAr, and ABr for the same interfacial Se arrangement, not AA?}
In these configurations, the Se sublattices facing each other across the interlayer region form an AB-type lateral registry. 
This local Se-interface motif avoids direct Se-on-Se vertical alignment and was therefore expected to provide favorable binding environments. 
The comparison in Fig.~\ref{fig:bilayer_DFT}(b) tests whether DFT assigns a clear energetic hierarchy within this structurally motivated subset of locally favorable stackings. 
Across the tested functionals, the relative energies of AB, AAr, and ABr span no more than 1.5 meV/f.u., making the three stackings effectively nearly degenerate at the DFT level. 
%\hscom{Same. It looks should be AB.}
The identity of the lowest-energy stacking also depends on the exchange-correlation functional: vdW-optB88, vdW-optB86b, rev-vdW-DF2, and vdW-DF2 predict ABr to be the lowest-energy stacking, whereas r${^2}$SCAN+rVV10, SCAN+rVV10, vdW-DF, PBE+MBD, and PBE+D3 favor AB.
The functional-dependent ordering therefore occurs within an extremely small energy window, and DFT does not provide a robust energetic hierarchy even within this common-Se-registry subset.
Interestingly, although AAr is connected to the experimentally reported $\beta$-like stacking motif, DFT does not isolate it as a distinct energetic minimum within the AB, AAr, and ABr subset. 
%\hscom{Same. It looks should be AB.}

Figure~\ref{fig:bilayer_DFT}(c) shows the corresponding optimized interlayer distances obtained from each DFT calculation.
%\hscom{Was the optimized distance obtained from full relaxation, or fitted result? I think it needs to be clarified}
The distance variation is more pronounced than the energy variation in Fig.~\ref{fig:bilayer_DFT}(b), indicating that DFT responds to lateral registry at the structural level even when the energetic hierarchy remains weak and depends on the functional.
This separation between structural and energetic corrugation is important for the present benchmark. 
A functional may predict different equilibrium distances for different stacking motifs, but that structural response does not by itself establish that the relative binding energies of those motifs are resolved with the same accuracy.

\begin{figure}
 \includegraphics[width=6.5in]{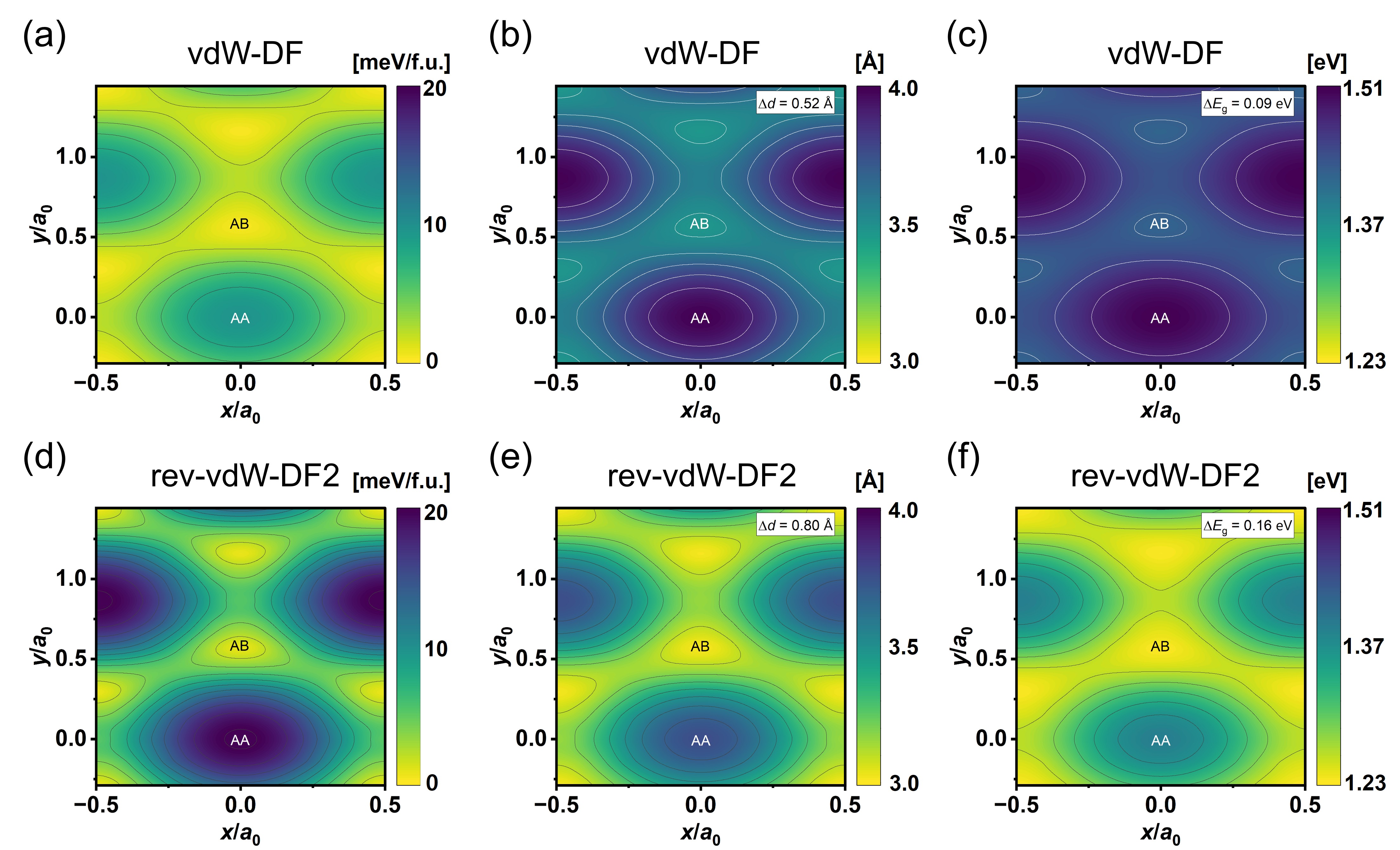}
 \caption{DFT sliding landscapes of bilayer InSe. Stacking-dependent energy maps ((a) and (d)), interlayer distances ((b) and (e)), and band gaps ((c) and (f)) obtained by laterally translating one InSe layer relative to the other using vdW-DF (upper panels) and rev-vdW-DF2 (lower panels).
 }
 \label{fig:DFT_sliding}
\end{figure} 

The functional spread in Fig.~\ref{fig:bilayer_DFT} sets the choice of representative DFT sliding landscapes. 
vdW-DF gives one of the weakest energetic separations among the selected high-symmetry stackings, while rev-vdW-DF2 gives the largest one. 
We therefore use these two functionals in Fig.~\ref{fig:DFT_sliding} as approximate lower and upper DFT estimates of the local-registry landscape. 
This extension from selected stackings to the full lateral-displacement space is required for twisted bilayers because a moir\'e pattern samples both favorable and unfavorable local registries.

Figure~\ref{fig:DFT_sliding} shows that this functional dependence extends across the full sliding landscape, not only at the high-symmetry points. 
On the common energy scale in Figs.~\ref{fig:DFT_sliding}(a) and \ref{fig:DFT_sliding}(d), vdW-DF gives a weakly corrugated energy landscape, whereas rev-vdW-DF2 gives a much stronger separation between low- and high-energy regions. 
The same contrast appears in the absolute interlayer distances and band gaps. 
vdW-DF predicts larger interlayer separations throughout the sliding landscape, and these larger separations are accompanied by larger band gaps. 
In contrast, rev-vdW-DF2 predicts shorter interlayer separations and smaller band gaps. 
This trend is consistent with the \(p_z\)-mediated interlayer-hybridization picture reported for PtTe$_{2}$~\cite{han2022strong,ahn2024exploring} and PtSe$_{2}$~\cite{li2021layer,ahn2025stacking}: reducing the interlayer distance enhances the out-of-plane overlap between the interfacial Se-derived density tails, which modifies the band-edge states and narrows the gap (See Supplementary Fig. 1). 
Conversely, the larger vdW-DF separations weaken this interlayer hybridization and keep the gap larger.

The magnitude of the registry dependence follows the same pattern. 
The distance corrugation increases from $\Delta d=0.52$~\AA\ in vdW-DF to $\Delta d=0.80$~\AA\ in rev-vdW-DF2, and the band-gap variation increases from $\Delta E_g=0.09$~eV to $\Delta E_g=0.16$~eV. 
Thus, the functional that predicts a stronger high-symmetry energy separation in Fig.~\ref{fig:bilayer_DFT} also predicts stronger structural and electronic registry dependence over the continuous sliding coordinate. 
The two DFT descriptions therefore imply qualitatively different local-registry landscapes for twisted bilayer InSe.

This difference is central to the need for a many-body benchmark. A twisted bilayer inherits its local relaxation, registry-energy cost, and spatial band-gap modulation from the same sliding landscape. 
The nearly flat vdW-DF landscape and the more strongly corrugated rev-vdW-DF2 landscape would lead to different descriptions of moir'e-scale interlayer response. 
The DFT local-registry landscape is therefore not a settled reference for twisted bilayer InSe. 
The DMC calculations below test which parts of this functional-dependent picture survive when the interlayer binding is treated at the many-body level.

\subsection*{Many-body monolayer reference}

\begin{table}[t]
\centering
\caption{
In-plane lattice constant \(a\), vertical In-In distance \(z_1\), and vertical Se-Se distance \(z_2\) for monolayer InSe. All values are given in \AA.
}
\label{tab:monolayer-inse-structure}
\begin{ruledtabular}
\begin{tabular}{lccc}
Method & \(a\) & \(z_1\) & \(z_2\) \\
\hline
DMC & \(4.00(1)\) & \(2.82(1)\) & \(5.28(1)\) \\
Exp. (bulk) & \(4.002\)~\cite{mudd2013tuning}, \(4.005\)~\cite{politano2017indium} & -- & -- \\
%Exp. (bulk) & \(4.002~\cite{mudd2013tuning}, 4.005~\cite{politano2017indium}\) & -- & -- \\
\hline
vdW-optB88 & \(4.07\) & \(2.81\) & \(5.43\) \\
vdW-optB86b & \(4.04\) & \(2.80\) & \(5.42\) \\
rev-vdW-DF2 & \(4.05\) & \(2.80\) & \(5.42\) \\
vdW-DF2 & \(4.23\) & \(2.90\) & \(5.49\) \\
r$^{2}$SCAN+rVV10 & \(4.02\) & \(2.78\) & \(5.35\) \\
SCAN+rVV10 & \(4.02\) & \(2.76\) & \(5.32\) \\
vdW-DF & \(4.17\) & \(2.87\) & \(5.46\) \\
PBE+D3 & \(4.06\) & \(2.79\) & \(5.37\) \\
PBE+MBD & \(4.03\) & \(2.79\) & \(5.42\) \\
\end{tabular}
\end{ruledtabular}
\end{table}

Before benchmarking the bilayer local-registry energetics, we first define the isolated-layer reference used for the DMC binding energies. 
This reference is needed because the bilayer binding energy is measured relative to separated monolayers.
We optimized monolayer InSe at the DMC level using the in-plane lattice constant $a$, the vertical In--In separation $z_{1}$, and the vertical Se--Se separation $z_{2}$ as structural variables.

The DMC structural optimization was performed using the surrogate Hessian line-search method~\cite{shin2021optimized}. 
In this approach, the search direction is generated from an approximate Hessian, while the energy minimization is carried out on the DMC potential-energy surface using DMC total energies. 
We used the Hessian obtained from DFT-PBE calculations to define the search directions, because the DFT-PBE Hessian provides a cost-effective approximation for structural optimization. 
For the monolayer calculation, the DMC potential-energy surface was sampled using a 288-electron supercell at the $\Gamma$ point, and the reported structure was obtained after three line-search iterations.

The DMC optimization gives $a=4.00(1)$~\AA, $z_{1}=2.82(1)$~\AA, and $z_{2}=5.28(1)$~\AA. 
The in-plane lattice constant is consistent with the reported bulk values, $a=4.005$~\cite{politano2017indium} and $4.002$~\AA~\cite{mudd2013tuning}, within the statistical uncertainty. 
Most of the tested DFT functionals give monolayer geometries close to the DMC reference on the scale considered here, as shown in Table~\ref{tab:monolayer-inse-structure}. 
The main exceptions are vdW-DF and vdW-DF2, which predict noticeably larger $a$ and $z_{2}$. 
Thus, the monolayer benchmark establishes the many-body separated-layer reference for the binding-energy calculations and shows that the isolated-layer geometry entering the bilayer comparison is broadly consistent across most functionals, with vdW-DF and vdW-DF2 yielding more expanded monolayer structures.

\subsection*{DMC local-registry energy hierarchy}

\begin{figure}
 \includegraphics[width=4.5 in]{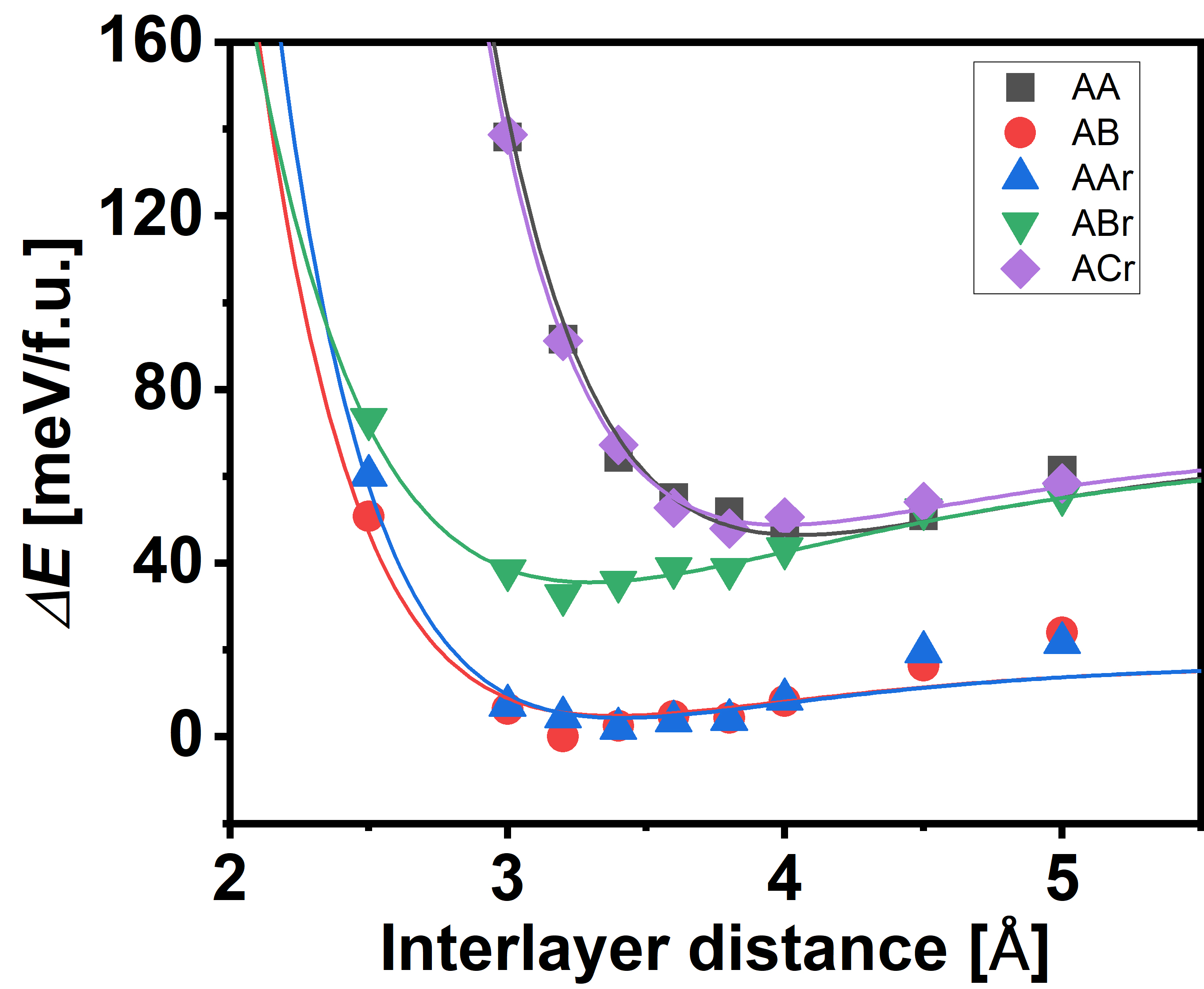}
 \caption{DMC binding-energy curves for the five bilayer InSe stackings. Dotted lines show Morse fits. Error bars are smaller than the symbols.} 
 \label{fig:bilayer_DMC}
\end{figure} 

\begin{table*}[t]
\centering
\caption{
Equilibrium interlayer distance $d$ and interlayer binding energy $E_b$ of InSe bilayers.
DMC values are obtained from the 128-atom supercell with finite-size-corrected binding energies.
DFT binding energies are computed relative to two isolated monolayers.
The units of $d$ and $E_b$ are \AA{} and meV/f.u., respectively.
}
\label{tab:DMC_DFT_binding}
\resizebox{\textwidth}{!}{
\begin{tabular}{c|cc|cc|cc|cc|cc}
\hline\hline
\multirow{2}{*}{Method}
& \multicolumn{2}{c|}{AA}
& \multicolumn{2}{c|}{AAr ($\beta$ phase)}
& \multicolumn{2}{c|}{AB ($\epsilon$ phase)}
& \multicolumn{2}{c|}{ABr}
& \multicolumn{2}{c}{ACr}
\\
& $d$ & $E_b$
& $d$ & $E_b$
& $d$ & $E_b$
& $d$ & $E_b$
& $d$ & $E_b$
\\
\hline
DMC
& 4.05(2) & -11(2)
& 3.41(7) & -70(3)
& 3.38(9) & -62(4)
& 3.30(3) & -29(3)
& 3.99(1) & -10(6)
\\
vdW-optB88
& 3.73 & -45.6
& 3.10 & -66.7
& 3.06 & -67.5
& 3.03 & -68.0
& 3.73 & -45.8
\\
vdW-optB86b
& 3.75 & -44.4
& 3.07 & -66.4
& 3.01 & -67.3
& 2.97 & -67.7
& 3.75 & -44.6
\\
rev-vdW-DF2
& 3.76 & -35.6
& 3.07 & -56.8
& 3.01 & -57.7
& 2.96 & -58.2
& 3.75 & -35.7
\\
r$^{2}$SCAN+rVV10
& 3.87 & -30.9
& 3.16 & -51.9
& 3.13 & -52.0
& 3.10 & -51.6
& 3.86 & -31.0
\\
SCAN+rVV10
& 3.87 & -27.0
& 3.19 & -46.4
& 3.11 & -46.6
& 3.09 & -46.2
& 3.86 & -27.2
\\
vdW-DF
& 3.99 & -38.3
& 3.49 & -48.9
& 3.46 & -49.0
& 3.46 & -49.0
& 3.99 & -38.3
\\
PBE+D3
& 3.78 & -40.8
& 3.12 & -60.8
& 3.09 & -61.3
& 3.07 & -61.1
& 3.78 & -40.9
\\
PBE+MBD
& 3.87 & -44.4
& 3.12 & -60.2
& 3.08 & -60.5
& 3.04 & -60.3
& 3.86 & -44.4
\\
\hline\hline
\end{tabular}
}
\end{table*}

We next benchmark the local-registry energetics of bilayer InSe at the many-body level. 
Figure~\ref{fig:bilayer_DMC} shows the DMC binding curves for the five high-symmetry bilayer registries considered in Fig.~\ref{fig:bilayer_DFT}. 
These curves were computed in a supercell containing 576 electrons and were used to determine the equilibrium interlayer distance for each registry. %\hscom{Number of electrons?}
The energetic hierarchy discussed below is obtained from the finite-size-corrected DMC binding energies evaluated at these optimized distances. 
This procedure provides a direct many-body reference for testing the shallow and functional-dependent DFT local-registry landscape established above.

The DMC equilibrium distances are broadly consistent with the structural expectations from the interfacial Se registry. 
AB, AAr, and ABr all relax to short interlayer separations, with $d=3.38(9)$, $3.41(7)$, and $3.30(3)$~\AA, respectively. 
Within the uncertainty of the optimized distances, these three stackings should be regarded as a comparable short-distance group rather than as clearly separated structures. 
In contrast, AA and ACr relax to much larger separations, $d=4.05(2)$ and $3.99(1)$~\AA, respectively. 
This is consistent with their less favorable interfacial geometry, where the facing Se sublattices are more directly aligned across the interlayer region. 
Thus, the structural relaxation mainly separates short-distance Se-registry motifs from long-distance, unfavorable Se-on-Se facing configurations.

The DMC interlayer binding energies strongly lift the DFT near-degeneracy within the common-Se-registry group.
First, the $\beta$-like AAr registry becomes the lowest-energy bilayer configuration, with $E_b=-70(3)$~meV/f.u. (See Table~\ref{tab:DMC_DFT_binding}).  
None of the tested DFT functionals in Fig.~\ref{fig:bilayer_DFT}(b) predicts AAr as the lowest-energy stacking. 
Second, the energetic separation between AAr and the competing $\epsilon$-like AB registry is enlarged at the many-body level.
In DFT, AAr and AB are weakly separated within the functional-dependent high-symmetry landscape, whereas DMC gives $E_b=-70(3)$ meV/f.u. for AAr and $E_b=-62(4)$ meV/f.u. for AB, an energy difference of 8(5) meV/f.u.. 
Third, ABr is no longer part of the near-degenerate low-energy group. 
Although ABr is close to AAr and AB in several DFT descriptions and is even selected by some functionals, DMC places it much higher in energy, with $E_b=-29(3)$ meV/f.u., corresponding to an energy difference of 41(4) meV/f.u. relative to AAr.
The many-body benchmark therefore does not simply reorder nearly degenerate DFT stackings. 
It expands a DFT energy window below 1.5 meV/f.u. into a separation of  41(4) meV/f.u., placing AAr and AB in the strongly bound group while clearly destabilizing ABr.

The ABr result is especially important because it shows that the short-distance Se-interface motifs are not energetically equivalent at the many-body level. 
AAr, AB, and ABr have comparable optimized interlayer distances within the DMC uncertainties, and the DFT relaxations also place them in the same short-distance structural group. 
Nevertheless, their DMC binding energies are strongly separated. AAr and AB remain strongly bound, whereas ABr is much less stable. 
Thus, the many-body hierarchy cannot be explained by the relaxed interlayer distance or by the coarse AB-type classification of the facing Se sublattices alone. 
The relevant distinction must come from the detailed lateral arrangement of the full interfacial environment.

The weakly bound AA and ACr stackings are also separated much more strongly from the low-energy group at the many-body level.
AA and ACr have binding energies of only -11(2) and -10(6) meV/f.u., respectively.
The resulting energy range between the most stable and least stable stackings is 60(7) meV/f.u. in DMC, compared with only 11–-23 meV/f.u. across the tested DFT functionals (see Table~\ref{tab:DMC_DFT_binding}).
This enlarged registry-energy range is directly relevant to twisted bilayer InSe.
For the $0^{\circ} \le \theta \le 30^{\circ}$ twist path considered below, AA provides the unfavorable local-registry limit relative to AB.
If DFT underestimates this AA-to-AB energy contrast, it will also underestimate the local-registry energy scale that drives moiré-scale relaxation.

The central outcome of the DMC benchmark is therefore not only that AAr is selected as the lowest-energy bilayer registry. 
It is that the many-body calculation reshapes the entire local-registry hierarchy relative to DFT. 
AAr and AB remain strongly bound, ABr loses the apparent low-energy stability suggested by DFT despite belonging to the same short-distance structural group, and AA and ACr become nearly unbound. 
This hierarchy motivates the interfacial charge-redistribution analysis below, which focuses on AAr, AB, and ABr to identify why ABr lacks the many-body stabilization present in the AAr and AB interfaces.

\subsection*{Registry-dependent density redistribution at the InSe interface}

%-----------------------------------------------------
\begin{figure}
 \includegraphics[width=6.5 in]{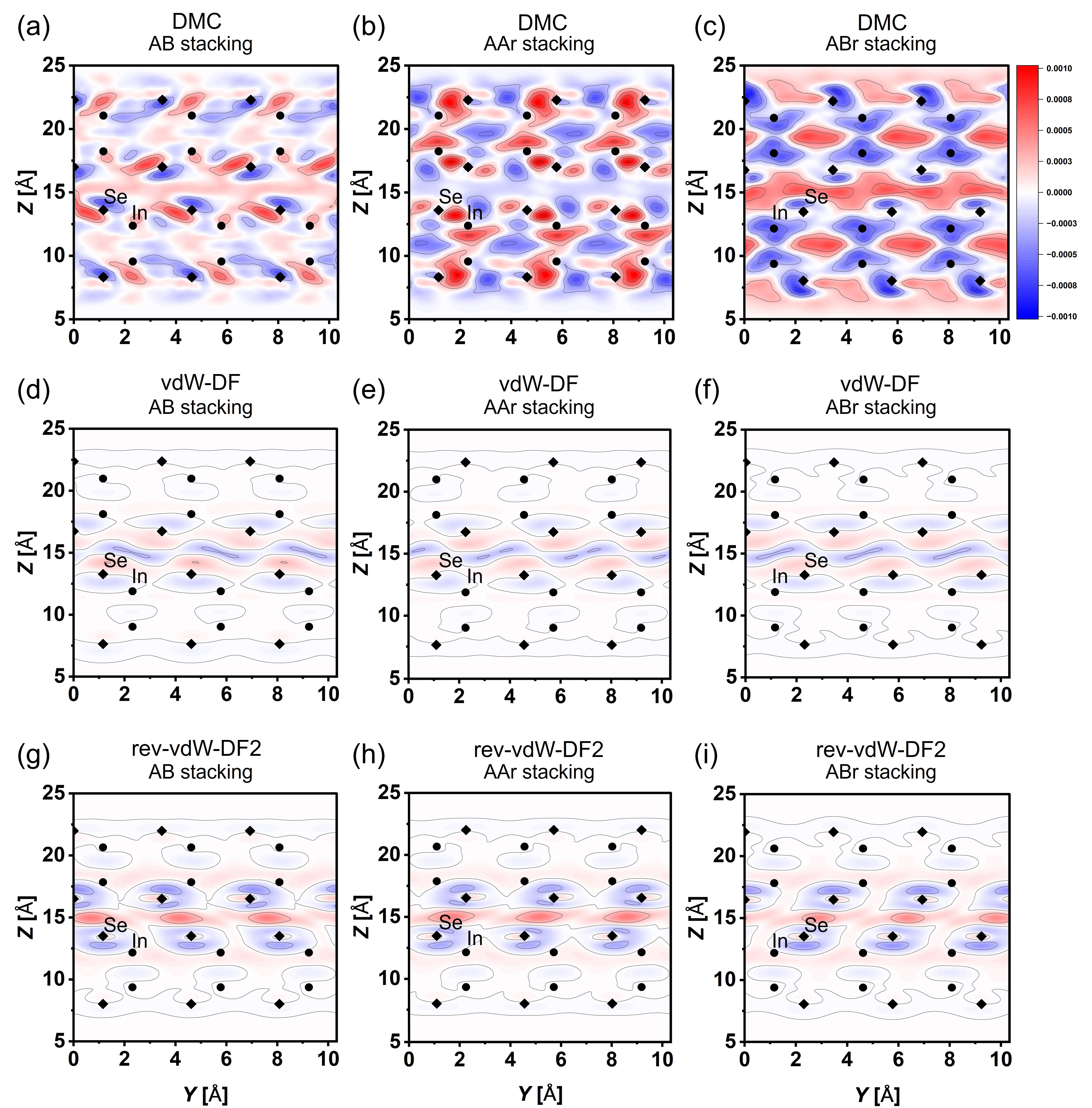}
 \caption{Registry-dependent charge redistribution upon bilayer formation, defined as \(\rho_{\mathrm{bilayer}}-\rho_{\mathrm{upper}}-\rho_{\mathrm{lower}}\), for the AB, AAr, and ABr stackings. Panels (a)--(c), (d)--(f), and (g)--(i) show the DMC, vdW-DF, and rev-vdW-DF2 results, respectively. All maps are shown in the \(yz\) plane. Red and blue indicate charge accumulation and depletion, respectively. Black circles and diamonds mark the In and Se positions. The color scale is in \(\AA^{-2}\).} 
 \label{fig:bilayer_density}
\end{figure}
%-----------------------------------------------------

To identify the microscopic origin of the DMC registry hierarchy, we analyze the charge redistribution induced by bilayer formation. We define the bilayer density difference as
$\Delta \rho = \rho_{\mathrm{bilayer}} - \rho_{\mathrm{upper}} - \rho_{\mathrm{lower}}$,
where the isolated upper- and lower-layer densities are evaluated in the same geometry as the bilayer. 
This definition removes the density of the separated monolayers and isolates the electronic response caused by bringing the two layers together. 
We focus on AB, AAr, and ABr because they form the short-distance structural group but are strongly separated in the DMC binding energies.

Figure~\ref{fig:bilayer_density}(a)--(c) shows that the DMC density response is strongly registry dependent even within the short-distance structural group. 
This point is clearest from the comparison between AAr and ABr. 
In AAr, bilayer formation depletes charge from the central interlayer region. 
This depletion is compensated by charge accumulation within the layers, especially around the In--Se bonding environment and the vertical In--In region. 
Thus, AAr responds to interlayer contact by removing charge from the overlap region between the two layers and redistributing it into the intralayer bonding network.

ABr shows the opposite response. 
In this registry, charge accumulates in the central interlayer region, while the nearby In--Se and In--In intralayer environments show depletion. 
This indicates that charge is redistributed from the intralayer bonding region toward the interlayer gap. 
The shorter interlayer distance of ABr therefore does not imply stronger binding. 
Instead, in this lateral registry, bringing the layers closer is accompanied by an unfavorable redistribution that enhances charge density in the interlayer region while depleting the intralayer bonding environment. 
This provides a natural explanation for why ABr loses the apparent low-energy stability suggested by DFT despite belonging to the same short-distance group as AAr and AB.

AB appears as a weaker version of the ABr-type response. 
It also shows charge accumulation in the interlayer region, but the redistribution is less intense than in ABr, and the depletion around the intralayer In--Se and In--In environments is correspondingly weaker. 
This weaker charge transfer is consistent with the DMC energetics: AB remains strongly bound, but it is less stable than AAr. 
The contrast between AAr and AB therefore shows that similar equilibrium distances can correspond to qualitatively different many-body density responses, while the contrast between AB and ABr shows that strengthening the interlayer accumulation pattern can reduce, rather than enhance, the net binding when it depletes the intralayer bonding network.

The DFT maps in Fig.~\ref{fig:bilayer_density}(d)--(i) do not reproduce this registry-specific contrast. 
Within each functional, the density rearrangements for AB, AAr, and ABr are much more similar to one another than in the DMC row. vdW-DF gives a weak and smooth redistribution near the interlayer region, while rev-vdW-DF2 gives a more localized pattern, but neither functional captures the reversal between AAr and ABr seen in DMC. 
As a result, DFT does not resolve the stacking-dependent charge transfer between the interlayer gap and the intralayer bonding network that separates the DMC binding energies of AAr, AB, and ABr.

The density analysis provides a microscopic explanation for the difference between the DMC hierarchy and the DFT landscape.
The enhanced stability of AAr is associated with a redistribution that removes charge from the interlayer overlap region and reinforces the intralayer bonding environment.
ABr shows the opposite redistribution, with charge transferred into the interlayer gap and depleted from the intralayer In--Se and In--In regions. 
AB follows the ABr-type response more weakly, which is consistent with its intermediate stability between AAr and ABr.
The many-body correction to the local-registry landscape is therefore not a uniform shift of the DFT energies. 
It changes the relative stability of the short-distance registries by resolving how each stacking redistributes charge between the interlayer gap and the intralayer bonding network.

\subsection*{Implications for twisted bilayer InSe}

%-----------------------------------------------------
\begin{figure}
 \includegraphics[width=6.5 in]{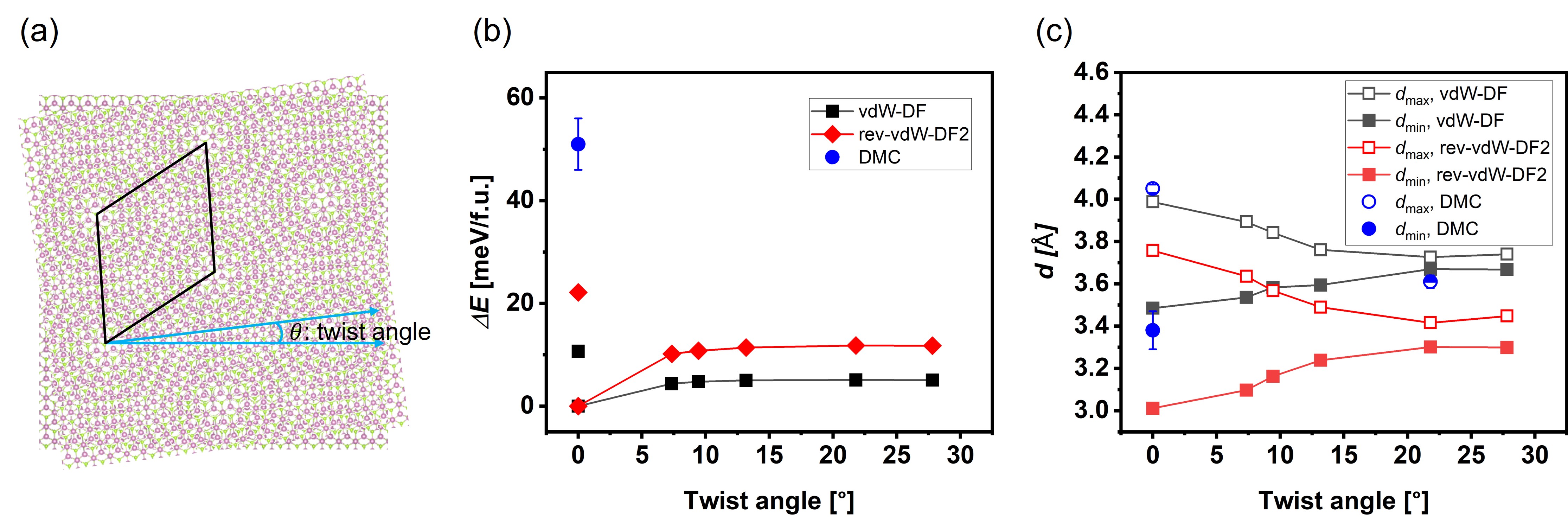}
 \caption{Twist-angle dependence from DFT and comparison with the DMC local-registry benchmark. (a) Structure of twisted bilayer InSe at a twist angle of 7.34$^{\circ}$. (b) Relative total energy of twisted bilayer InSe as a function of twist angle from vdW-DF and rev-vdW-DF2, compared with the DMC energy scale for the high-symmetry local-registry benchmark. (c) Twist-angle dependence of the minimum and maximum interlayer distances obtained from DFT relaxation, compared with the DMC equilibrium distances.%\hscom{Explanation is needed for blue half-solid circle at 22.5$^\circ$. If solid lines are just guide for the eyes, why isn't there the line from 0 $^\circ$ to 22.5 $^\circ$ for DMC? Are all they independent?}
 }
 \label{fig:twisted}
\end{figure}
%-----------------------------------------------------

We now connect the local-registry benchmark to twisted bilayer InSe. 
As illustrated by the twisted structure shown in Fig.~\ref{fig:twisted}(a), the twist angle $\theta$ is defined by rotating one InSe layer relative to the other away from the AB-stacked bilayer reference configuration. %\hscom{Do we have a simple top view of twisted InSe bilayer? It would be better to visualize how $\theta$ is defined.} 
In the $0^\circ \leq \theta \leq 30^\circ$ range considered here, the moir\'e pattern samples local environments related to AA, AB, and AC stackings. %\hscom{What is AC stacking? Was it mentioned or defined previously in this manuscript?}
Because AB and AC are equivalent stacking configurations in bilayer InSe, the relevant local registry contrast is between the favorable AB-like regions and the unfavorable AA-like regions. 
Accordingly, the two points at $\theta=0^\circ$ in Fig.~\ref{fig:twisted}(b) mark these two high-symmetry limits: the lower point is the AB reference, set to zero in relative energy, and the upper point is the unstable AA stacking.

Figure~\ref{fig:twisted}(b) shows that the DFT twist-angle energetics follow this local-registry picture. 
The energies of the twisted structures lie between the AB and AA limits, consistent with the view that a twisted bilayer samples a spatial mixture of local registries rather than forming a new uniform stacking configuration. 
In this sense, the twist energy behaves like a moir\'e-scale average over the underlying local-registry landscape. 
The difference between the two DFT descriptions also follows the trend established in the sliding maps. 
vdW-DF gives a weak local energy scale and a shallow twist-angle dependence, whereas rev-vdW-DF2 gives a larger AB-to-AA separation and a stronger twist-angle dependence. 
Thus, the DFT twist energetics directly inherits the local-registry energy scale seen in Fig.~\ref{fig:DFT_sliding}, so any compression of that scale propagates into the predicted moir'e energetics.

The DMC point at $\theta=0^\circ$ gives the many-body AA-to-AB energy difference on the same relative-energy scale. 
This DMC energy separation is much larger than the corresponding DFT estimates. 
Since the twisted-bilayer energy is governed by the local-registry window defined by the AB and AA limits, the larger DMC separation shows that DFT underestimates the registry-energy scale entering the twist problem.
This underestimation should be most important at smaller twist angles, where the moir'e length scale is large and the structure can form well-defined local stacking patches. 
In that regime, the twist-angle energy should be more sensitive to the many-body registry contrast than suggested by the DFT curves.
The limitation of the DFT-based moir\'e description is therefore already apparent from the local-registry DMC benchmark.

Figure~\ref{fig:twisted}(c) shows the corresponding structural trend. 
The DFT-relaxed twisted bilayers develop a range of local interlayer distances, bounded by $d_{\min}$ and $d_{\max}$.
These values reflect local moir'e corrugation: AB-like regions relax toward shorter distances, whereas AA-like regions relax toward larger distances. 
At smaller twist angles, the moir\'e length scale is large compared with the lattice constant, so the local stacking patches are more distinct and the separation between $d_{\min}$ and $d_{\max}$ becomes clearer. 
At larger twist angles, the moir\'e length scale approaches the lattice scale, the local-patch description becomes less distinct, and $d_{\min}$ and $d_{\max}$ move closer together.

The two functionals give different structural corrugations as displayed in Fig.~\ref{fig:twisted}(c). 
The DFT-relaxed twisted structures develop corrugated layers with distinct local $d_{\min}$ and $d_{\max}$ values, shown by the filled and open symbols, respectively. 
rev-vdW-DF2 predicts shorter interlayer distances and a larger separation between $d_{\min}$ and $d_{\max}$, whereas vdW-DF gives larger distances and a shallower corrugation. 
For the high-symmetry bilayer registries, the DMC equilibrium distances are closer to the vdW-DF distance range than to the shorter rev-vdW-DF2 range. 
For the \(21.79^\circ\) twisted bilayer, the blue half-filled circle denotes the DMC equilibrium separation obtained for two rigid, flat monolayers. 
This effective rigid-layer separation falls between the vdW-DF \(d_{\min}\) and \(d_{\max}\) values at the same twist angle, suggesting that vdW-DF provides a reasonable estimate of the local interlayer-distance range. 
However, this structural agreement does not validate the DFT local-registry energetics, which remain much more weakly corrugated than the DMC benchmark.
%The average-like DMC distance falls between the vdW-DF $d_{\min}$ and $d_{\max}$ values at the same twist angle, suggesting that vdW-DF provides a reasonable estimate of the local interlayer-distance range.
However, this structural agreement does not validate the DFT local-registry energetics, which remain much more weakly corrugated than the DMC benchmark.

The main implication is that DFT-based predictions for twisted bilayer InSe can miss the many-body local-registry physics even when the relaxed geometry appears reasonable. 
The DMC benchmark shows that the energetic contrast between favorable and unfavorable local registries is much larger than in DFT. 
For the $0^{\circ} \le \theta \le 30^{\circ}$ twist path considered here, the directly relevant benchmark is the much larger DMC energy contrast between the AB- and AA-like limits, which shows that DFT underestimates the registry-energy scale driving moir\'e relaxation. 
More generally, the AAr, AB, and ABr comparison shows that DFT can also render distinct atomic registries nearly degenerate when they share the same facing-Se arrangement, even though DMC separates them by tens of meV/f.u.. 
DFT-based moir\'e models can therefore inherit both an underestimated contrast between favorable and unfavorable local environments and artificial near-degeneracies among locally similar registries.
The weak twist dependence found in DFT should therefore not be interpreted as intrinsically weak registry dependence, but as a consequence of a flattened local-registry landscape. 
The bilayer benchmark thus exposes a central limitation of DFT-based moir\'e modeling: even when the relaxed geometry appears reasonable, the energetic driving forces for structural reconstruction, domain formation, and local electronic reconstruction can be substantially underestimated. 
More generally, reliable moir\'e modeling requires validating the local-registry energy hierarchy itself, not only the relaxed geometry.

\section*{Conclusion}
\label{sec:conclusion}

We have benchmarked the local-registry energetics of bilayer InSe using DMC. 
DFT predicts a shallow and functional-dependent registry landscape in which AB, AAr, and ABr remain nearly degenerate within 1.5 meV/f.u., although they differ in their full atomic registries, and the selected minimum depends on the exchange-correlation functional.
This behavior persists when the high-symmetry comparison is extended to representative lower and upper DFT estimates of the full sliding landscape. 
DFT captures registry-dependent structural relaxation, but it does not provide a robust energetic hierarchy or energy scale for the local stacking environments relevant to twisted bilayer InSe.

The DMC benchmark gives a qualitatively different hierarchy and energy scale. 
DMC lifts the apparent degeneracy among AAr, AB, and ABr, separating them by 8(5) and 41(4) meV/f.u. relative to AAr, and expands the energy difference between the most stable and least stable registries to 60(7) meV/f.u..
The charge-redistribution analysis shows that this hierarchy is associated with registry-specific rearrangement between the interlayer gap and the intralayer bonding network, which is largely smoothed out in DFT. 
For the twist path studied here, the enlarged DMC contrast between AB- and AA-like environments shows that DFT underestimates the registry-energy scale entering the moir\'e problem. 
The AAr, AB, and ABr comparison further demonstrates that DFT can artificially compress the energy differences among locally similar registries.
The weak twist dependence predicted by DFT should therefore be interpreted as a consequence of this flattened registry landscape, which can underestimate the driving forces for structural relaxation, domain formation, and local electronic reconstruction.
More broadly, the DMC results show that reliable moir\'e modeling requires validation of the local-registry hierarchy and its full energetic scale, not merely the relaxed geometry or the identity of the DFT minimum.

\section*{Methods}
\label{sec:methodology}

%We used DMC within the fixed-node approximation as implemented in the QMCPACK code~\cite{QMCPACK}. Single Slater-determinant wavefunctions were used as trial wavefunctions in the QMC algorithm, with up to three-body Jastrow correlation coefficients in order to incorporate electron-ion, electron-electron, and electron-electron-ion correlations. Cut-offs for the one- and two-body Jastrows were set as the Wigner-Seitz radius of the given supercell while a maximum of 5.0~Bohr was used as the cut-off for the short-range three-body term. Single-particle orbitals in the QMC trial wavefunctions were generated by solving the Kohn-Sham equations using DFT. The DFT calculations used to generate the single-particle orbitals for QMC were performed using the QUANTUM ESPRESSO code~\cite{giannozzi09}, with a plane-wave kinetic-energy cutoff of \(400\ \mathrm{Ry}\) and an \(8\times8\times1\) Monkhorst--Pack \(k\)-point grid. Kohn-Sham orbitals in the Slater determinant were generated using PBE parametrization~\cite{perdew96} of the generalized gradient approximation exchange-correlation functional. Energy-consistent norm-conserving pseudopotentials proposed by Burkatzki, Filippi and Dolg were used for the In and Se atoms in DFT and DMC calculations~\cite{burkatzki07}. DMC calculations were performed using a time step of 0.005~Ha$^{-1}$ within the non-local $T$-move approximation~\cite{casula10}.
We used DMC within the fixed-node approximation as implemented in QMCPACK~\cite{QMCPACK}. 
Single-determinant Slater--Jastrow trial wave functions were used, including one-body electron--ion, two-body electron--electron, and three-body electron--electron--ion Jastrow terms.
The cutoffs of the one- and two-body Jastrow terms were set to the Wigner--Seitz radius of each supercell, while the three-body cutoff was limited to \(5.0\) bohr. 
The single-particle orbitals used in the Slater determinants were generated from PBE~\cite{perdew96} calculations performed using QUANTUM ESPRESSO~\cite{giannozzi09}, with a plane-wave kinetic-energy cutoff of \(400\ \mathrm{Ry}\) and an \(8\times8\times1\) Monkhorst--Pack \(k\)-point grid. 
Energy-consistent norm-conserving pseudopotentials proposed by Burkatzki, Filippi, and Dolg were used for both In and Se in the DFT and DMC calculations~\cite{burkatzki07}. 
DMC calculations were performed with a time step of \(0.005\ \mathrm{Ha}^{-1}\) using the nonlocal T-move approximation~\cite{casula10}.

Because the density operator does not commute with the Hamiltonian, its standard DMC expectation value is a mixed estimator with a leading-order trial-wave-function bias. We therefore used the extrapolated estimator, $\rho_{\text{ext}}(r)=2\rho_{\text{DMC}}(r)-\rho_{\text{VMC}}(r)$, which cancels the leading trial-wave-function bias and leaves an error that is second order in the difference between the trial and fixed-node wave functions~\cite{foulkes01}.
To reduce one-body finite-size effects, we employed twist-averaged boundary conditions~\cite{lin01} with up to 64 twists for the InSe cells.
We further evaluated twist-averaged DMC energies for three supercell sizes and extrapolated them to the thermodynamic limit to reduce two-body finite-size effects (See Supplementary Fig. 3).  
The DMC calculation for the 21.79$^{\circ}$ twisted bilayer was performed using a 28-f.u. supercell containing 504 electrons. Its interlayer binding curve was obtained by rigidly varying the separation between two flat monolayers without allowing intralayer relaxation or out-of-plane corrugation. 
The resulting equilibrium distance therefore represents an effective rigid-layer separation rather than either of the local \(d_{\min}\) and \(d_{\max}\) values of a corrugated moiré structure.

Stacking-dependent DFT calculations for the high-symmetry bilayers and sliding landscapes were performed using the projector-augmented-wave method as implemented in the VASP code~\cite{kresse1993ab,kresse1996efficient,kresse1999ultrasoft}.
The plane-wave kinetic-energy cutoff was set to 400 eV, and Brillouin-zone integrations were performed using $8 \times 8 \times 1$ Monkhorst-Pack $k$-point meshes for the primitive bilayer cells. 
The atomic positions and interlayer distance were relaxed until the residual forces were smaller than 10$^{-3}$ eV~\AA$^{-1}$.
To assess the functional dependence of the local-registry landscape, we compared a set of nonlocal and dispersion-corrected exchange-correlation functionals, including vdW-DF~\cite{vdw-df}, vdW-DF2~\cite{vdW-DF2}, rev-vdW-DF2~\cite{rev-vdW-DF2}, vdW-optB88~\cite{optB88}, vdW-optB86b~\cite{optB86b}, SCAN+rVV10~\cite{scan_rvv10}, r$^{2}$SCAN+rVV10~\cite{r2scan_rvv10}, PBE+D3~\cite{pbed3} and PBE+MBD~\cite{mbd}.
These calculations provide the DFT reference for the relative stacking energies, equilibrium interlayer distances, band gaps, and relaxed twist-angle structures compared with the DMC local-registry benchmark.

\section*{Data Availability}
The data that support the findings of this study are available
this article, its supplementary material, and the materials
data facility\cite{Blaiszik2016, Blaiszik2019} at [link to be provided upon acceptance of this manuscript].

\section*{Acknowledgements}
J. Ahn acknowledges support from U.S. Department of Energy, Office of Science, Office of Basic Energy Sciences, Computational Materials Sciences Award No. DE-SC0020177 for calculations, analysis, and writing of the paper. 
A.A. (analysis, pseudopotentials) and H.S. (mentorship, calculations, analysis, writing) was supported by the U.S. Department of Energy, Office of Science, Basic Energy Sciences, Materials Sciences and Engineering Division, as part of the Computational Materials Sciences Program and Center for Predictive Simulation of Functional Materials. 
J.N. (analysis, writing) and N. P. S. (mentorship, analysis, writing) acknowledge support from the National Science Foundation under Award No. DMR-1905986.
An award of computer time was provided by the Innovative and Novel Computational Impact on Theory and Experiment (INCITE) program. This research used resources of the Argonne Leadership Computing Facility, which is a DOE Office of Science User Facility supported under contract DE-AC02-06CH11357.

\section*{References}
\bibliography{main.bib}
\bibliographystyle{naturemag}
\end{document}